\documentclass[aps,amsfonts,amsmath,amssymb,nofootinbib,twocolumn,superscriptaddress]{revtex4}

\usepackage{braket}
\usepackage{graphics}
\usepackage{hyperref}
\usepackage{epsfig}
\usepackage{color}

\usepackage[normalem]{ulem}
\usepackage{multirow}
\usepackage{bm}

\usepackage{pifont}
\usepackage{amssymb}
\usepackage{amsmath}

\def\ket#1{|#1\rangle }
\def\bra#1{\langle #1 |}

\begin{document}

\bibliographystyle{naturemag}

\title{Three-state nematicity and magneto-optical Kerr effect in the charge density waves in kagome superconductors}

\author{Yishuai Xu}
\affiliation{Department of Physics and Astronomy, University of Pennsylvania, Philadelphia, Pennsylvania 19104, U.S.A}
\author{Zhuoliang Ni}
\affiliation{Department of Physics and Astronomy, University of Pennsylvania, Philadelphia, Pennsylvania 19104, U.S.A}
\author{Yizhou Liu}
\affiliation{Department of Condensed Matter Physics, Weizmann Institute of Science, Rehovot, Israel}
\author{Brenden R. Ortiz}
\affiliation{Materials Department and California Nanosystems Institute, University of California Santa Barbara, Santa Barbara, California
93106, USA}
\author{Qinwen Deng}
\affiliation{Department of Physics and Astronomy, University of Pennsylvania, Philadelphia, Pennsylvania 19104, U.S.A}
\author{Stephen D. Wilson}
\affiliation{Materials Department and California Nanosystems Institute, University of California Santa Barbara, Santa Barbara, California
93106, USA}
\author{Binghai Yan}
\affiliation{Department of Condensed Matter Physics, Weizmann Institute of Science, Rehovot, Israel}
\author{Leon Balents}
\affiliation{Kavli Institute for Theoretical Physics, University of California, Santa Barbara, Santa Barbara,
California 93106, USA}
\author{Liang Wu}
\email{liangwu@sas.upenn.edu}
\affiliation{Department of Physics and Astronomy, University of Pennsylvania, Philadelphia, Pennsylvania 19104, U.S.A}

\date{\today}

\begin{abstract}
The kagome lattice provides a fascinating playground to study geometrical frustration, topology and strong correlations. The newly-discovered kagome metals AV$_3$Sb$_5$ (where A can refer to either K, Rb, or Cs) exhibit phenomena including topological band structure, symmetry-breaking charge-density waves and superconductivity. Nevertheless, the nature of the symmetry breaking in the charge-density wave phase is not yet clear, despite the fact that it is crucial in order to understand whether the superconductivity is unconventional. In this work, we perform scanning birefringence microscopy on all three members of this family and find that six-fold rotation symmetry is broken at the onset of the charge-density wave transition in all compounds. We show that the three nematic domains are oriented 120$^\circ$ to each other and propose that staggered charge-density wave orders with a relative $\pi$ phase shift between layers is a possibility that can explain these observations. We also perform magneto-optical Kerr effect and circular dichroism measurements, and the onset of both signals is at the transition temperature, indicating broken time-reversal symmetry and the existence of the long-sought loop currents in that phase. 
\end{abstract}

\maketitle 

The kagome lattice has attracted tremendous research interest for decades as the corner sharing triangular lattice has inherent geometrical frustrations that  host exotic phases such as the quantum spin liquid state \cite{HanNature2012}. There has been  recent interest in the magnetic kagome systems from the perspective of the topological electronic structures such as the magnetic Weyl semimetals Mn$_3$Sn\cite{NakatsujiNature2015}, Co$_3$Sn$_2$S$_2$ \cite{belopolskiScience2019} and strongly correlated flat bands in Fe$_3$Sn$_2$ and FeSn \cite{YinNature2018,YinNatPhys2019,YinNature2020,YeNature2018}. Strong electronic correlations without magnetism can also lead to exotic phases such as  high-Tc superconductivity, but it has been often difficult to reveal different kinds of broken symmetries. Loop currents, originally proposed in cuprate superconductors \cite{VarmaPRB1997}, have also been predicted  in the kagome lattice \cite{ParkPRB2021, FengSciBulletin2021,LinPRB2021, DennerPRL2021}, but a clear evidence has been lacking. 

The newly discovered kagome metals AV$_3$Sb$_5$ (A=K, Rb, Cs) are recent examples of hosting charge density waves (CDWs) below $T_{CDW}$ $\approx$ 80-100 K and superconductivity below $T_{C}$ $\approx$ 0.9-2.5 K \cite{OrtizPRMat2019, OrtizPRL2020, OrtizPRMat2021, JiangNatMat2021, XuPRL2021, LiuPRX2021, ShumiyaPRB2021, WangPRB2021_STM, WangAM2021, LuoPRL2022, ChenNature2021, YinCPL2021}. Different from magnetic kagome materials, AV$_3$Sb$_5$ do not have a detectable local electronic moments \cite{MielkeNature2022}, but a surprisingly large anomalous Hall effect was reported with dominating extrinsic skew scattering mechanism\cite{yangSciAdv2020,YuPRB2021}. An increase  of the muon depolarization below the CDW transition temperature in zero-field $\mu$SR measurements in  KV$_3$Sb$_5$ and CsV$_3$Sb$_5$ have been interpreted as the evidence of the time-reversal symmetry (TRS) breaking, but the onset temperature is not always at $T_{CDW}$ and the muon depolarization is not directly related with the TRS-breaking order parameter \cite{MielkeNature2022}.  Therefore, a direct measurement of the TRS-breaking order  parameter at zero-field is urgently needed. Another major  debate in the community is whether  the system   has a six-fold or two-fold rotational symmetry in the CDW phase \cite{WuarXiv2021, OrtizPRX2021, ZhaoNature2021,XiangNatComm2021,NiCPL2021,NieNature2022,LiNatPhys2022}, and at what temperature  the six-fold symmetry breaks. Almost all of the experiments that claimed the two-fold symmetry were  performed at temperature much below the CDW transition\cite{ZhaoNature2021,NieNature2022,LiNatPhys2022}. Therefore, whether the two-fold symmetry is directly related to the CDW has not been clear.  In this work, we use scanning birefringence microscopy, magneto-optical Kerr effect, and circular dichroism  to reveal that the CDW transition temperature is the onset of six-fold rotational symmetry breaking and TRS breaking. Our micron-scale imaging bridges the gap between nano-scale scanning probes and macroscopic measurements, and provides new insight and strong constrains on the interpretation of many results by macroscopic probes.  \\

\textbf{Three-state nematicity}

\begin{figure*}
    \centering
    \includegraphics[width=\textwidth]{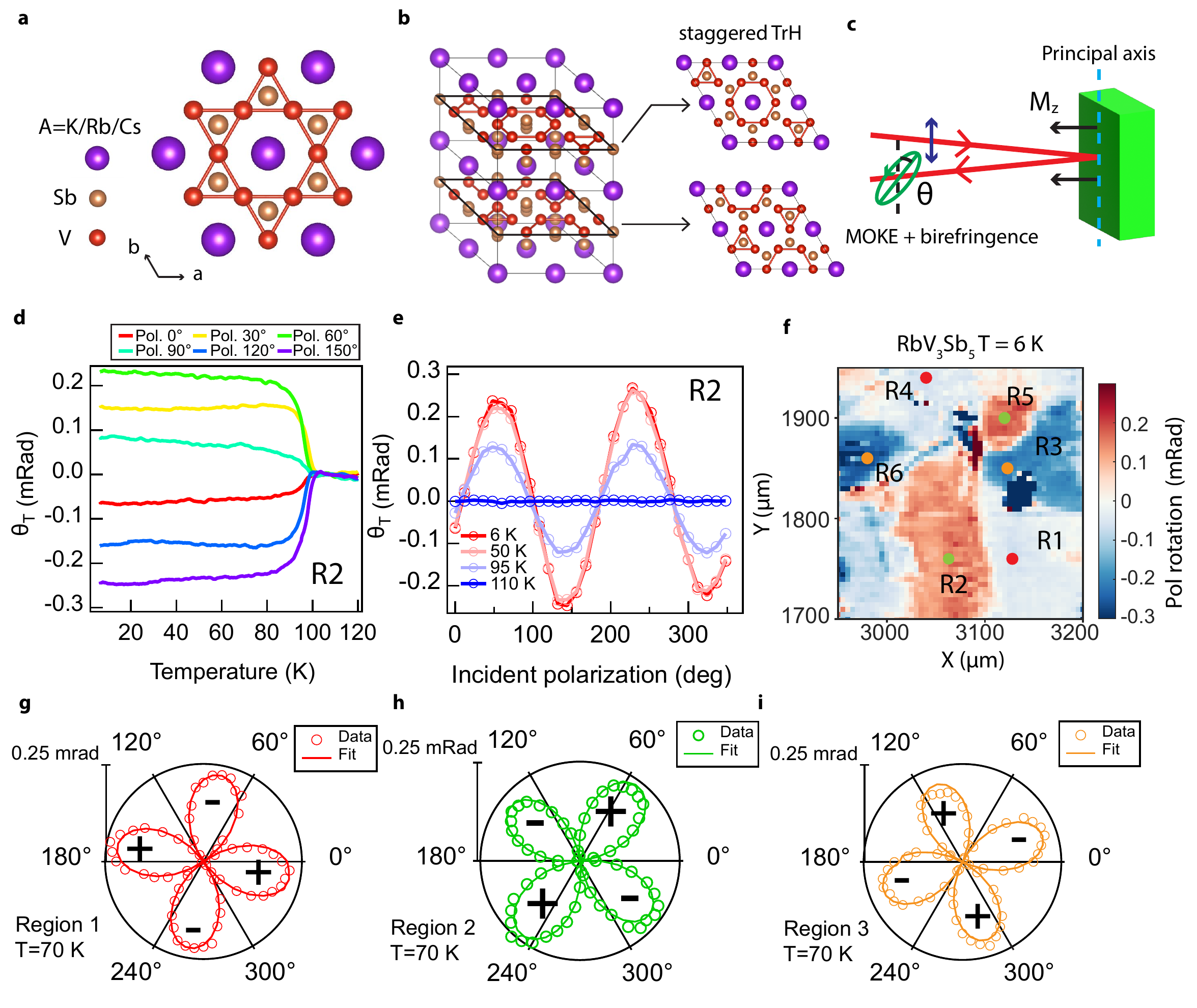}
    \caption{\textbf{Three-state nematic order in RbV$_3$Sb$_5$.} (a) Top view of the kagome lattice structure (a-b plane) of AV$_3$Sb$_5$ (A=K, Rb, Cs). The V$_3$Sb and K/Rb/Cs layers form the kagome and triangular lattices, respectively. (b) A 3D crystal structure showing an staggered tri-hexagonal  (TrH) CDW order. There is a relative phase of $\pi$ shift between the neighboring in-plane 2$\times$2 tri-hexgonal CDW orders. (c) A schematic shows the change of polarization angle $\theta_{T}$ due to a combination of magneto-optical Kerr effect and the birefringence effect. (d) $\theta_T$ vs temperature measured at the green spot in region 2 in (f) with different incident polarization. (e) $\theta_T$ vs incident polarization at different temperature. The plot is obtained from averaging the data shown in (d) in a $\pm$1 K window centered around the labeled temperature. (f) Spatial mapping of $\theta_T$ at T= 6 K in a RbV$_3$Sb$_5$ sample measured at an incident polarization of $\phi$=37.6$^\circ$, which corresponds to zero birefringence in R1. Six regions are labeled as R1 to R6, and the red/green/orange spots indicate the positions where the birefringence measurement within each domain are performed. The two dark blue islands are caused by impurities on the surface (see optical images in Extended data Fig.\ref{extfig1}). (g-i) Polar plots of the birefringence patterns at T= 70 K at the marked red/green/orange dots in regions 1-3, respectively. The + and - symbols show the sign of $\theta_T$. }
    \label{fig:fig1}
\end{figure*}

AV$_3$Sb$_5$ (A=K, Rb, Cs) share a hexagonal crystal structure, consisting of a kagome lattice of V atoms coordinated by Sb in the V-Sb sheet stacked between the A sheets, as shown in Fig. \ref{fig:fig1}(a). Therefore, it is six-fold rotationally symmetric in the normal state above $T_{CDW}$. In the CDW phase, the 2$\times$2 superlattice per layer could form the star-of-David  or the tri-hexgonal pattern, which still keep the six-fold symmetry in the pristine lattice. Nevertheless, a $\pi$ phase shift between neighboring layers can reduce the symmetry to two-fold, as shown in Fig. \ref{fig:fig1}(b) and Extended data Fig.\ref{extfig1}(a). Note that electronically-driven nematicity is also two-fold symmetric, but the onset temperature is much below $T_{CDW}$ as observed by scanning tunneling microscopy (STM)\cite{ZhaoNature2021,NieNature2022,LiNatPhys2022}. 

To study the rotational symmetry in the CDW phase, we perform scanning birefringence measurements as shown in Fig. \ref{fig:fig1}(c).  Under normal incidence, the change of the polarization, $\theta_T$, depends not only on the out-of-plane magnetization or orbital moment, known as the polar magneto-optical Kerr effect (MOKE), but also on the birefringence term when the rotational symmetry is lower than three-fold. We can distinguish two contributions by rotating the polarization of the incident beam,
\begin{equation}
    \theta_T = \theta_K + \theta_B  \mathrm{sin} (2\phi-\phi_0) \label{eq:eq1}
\end{equation}
where  $\theta_K$ and $\theta_B$ represent the real part of the MOKE and the amplitude of the birefringence; $\phi$ is the polarization angle of the incident light with respect to the horizontal axis in the lab. $\phi_0$ is one principal axis of the crystal in the lab frame. Note that the ``MOKE", $\theta_K+i\eta$, is actually a complex quantity, and the $\theta_K$ and $\eta$ are often called MOKE and ellipticity, respectively. (See Methods.) Fig. \ref{fig:fig1}(d) shows temperature-dependent $\theta_T$ at different polarization for RbV$_3$Sb$_5$, and the onset of $\theta_T$ is at  $T_{CDW}$ $\approx$ 103 K. As we change the polarization of the incident light, both the sign and the magnitude of $\theta_T$ changes with a maximum amplitude of 0.27 mRad at 6 K. Plotting the $\theta_T$ vs the incident polarization at a constant temperature shows a two-fold symmetric pattern below  $T_{CDW}$ (see Fig. \ref{fig:fig1}(e)). In contrast, we barely observe any angle dependence above $T_{CDW}$, which is consistent with the six-fold symmetry of the kagome lattice.   The two-fold symmetric pattern originates from the breaking of the six-fold rotation symmetry due to the formation of the CDW.

\begin{figure*}
    \centering
    \includegraphics[width=0.8\textwidth]{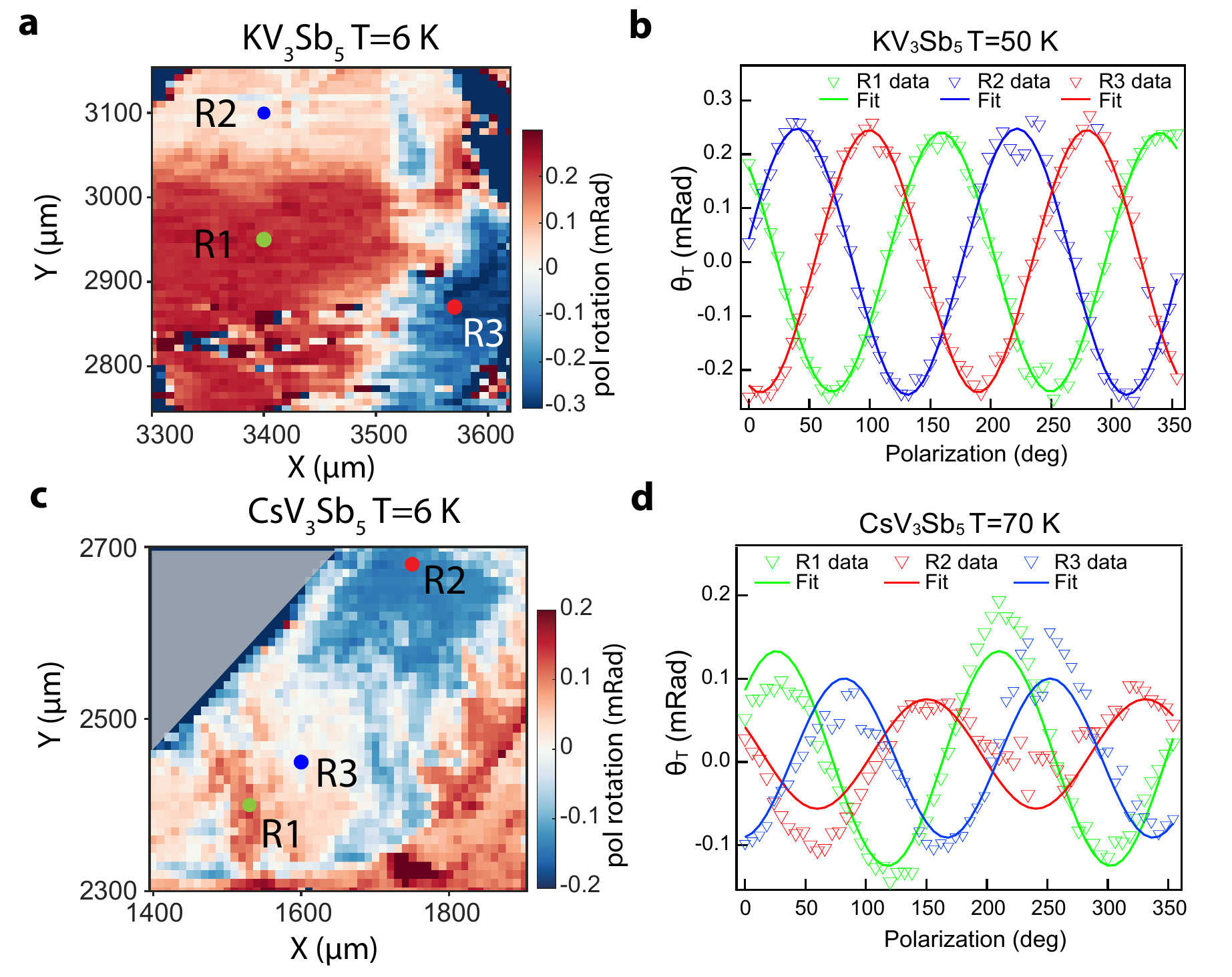}
    \caption{\textbf{Three-state nematic order in KV$_3$Sb$_5$ and CsV$_3$Sb$_5$.} (a, c) Spatial mapping of $\theta_T$ at T= 6 K in KV$_3$Sb$_5$ and CsV$_3$Sb$_5$ measured at incident polarizations of 0$^\circ$ and 46$^\circ$, respectively. Three domains are labelled as R1, R2 and R3. The red/green/blue spots indicate the positions where the birefringence measurements (see panels b and d) are performed. The gray triangle in panel (c) indicates region off the sample. (b, d) Birefringence patterns of KV$_3$Sb$_5$ and CsV$_3$Sb$_5$ measured at T= 50 K and T= 70 K at green/red/blue dots in regions 1-3, respectively.}
    \label{fig:fig2}
\end{figure*}

We set the incident polarization at a fixed polarization, and perform a  mapping of $\theta_T$ with a spatial resolution of 8 $\mu$m as shown in Fig. \ref{fig:fig1}(f). Three distinct domains marked with red, white and blue are clearly seen in the map. An angle dependent birefringence measurements in the selected six regions show that the regions with the same color have the same polar patterns, as shown in Fig. \ref{fig:fig1}(g-i) and extended Fig.1 \ref{extfig1}(e-g). Between regions with different color, the polar patterns are rotated by approximately 120$^\circ$ to each other. Fig. 1f shows a possible topological defect as observed in YMnO$_3$ \cite{Choi_NatMat_2010}. However, a surface defect exists where the six regions meet, which prevent a clear conclusion. (see the optical image in Extended Fig.\ref{extfig1}(b)). Fig. \ref{fig:fig2} shows that KV$_3$Sb$_5$ and CsV$_3$Sb$_5$ also have three domains, where the principal axis is also rotated by 120$^\circ$ from each other. A relatively larger deviation from the fitted curve is observed for the Cs compound, which might be due to the slightly uneven surface resulted from cleaving (see Extended data Fig.\ref{extfig2}(b-c)). The temperature dependence of the birefringence of KV$_3$Sb$_5$ and CsV$_3$Sb$_5$ also shows that the onset of  six-fold rotational symmetry breaking is  at $T_{CDW}$, $\approx$ 74 K and $\approx$ 92 K for these two compounds respectively (See Extended data Fig.\ref{extfig3}). To summarize, the three nematic domains  are a universal feature of the CDW phases in AV$_3$Sb$_5$(A=Cs, Rb, K).\\

\textbf{Polar Magneto-optical Kerr effect}

\begin{figure*}
    \centering
    \includegraphics[width=\textwidth]{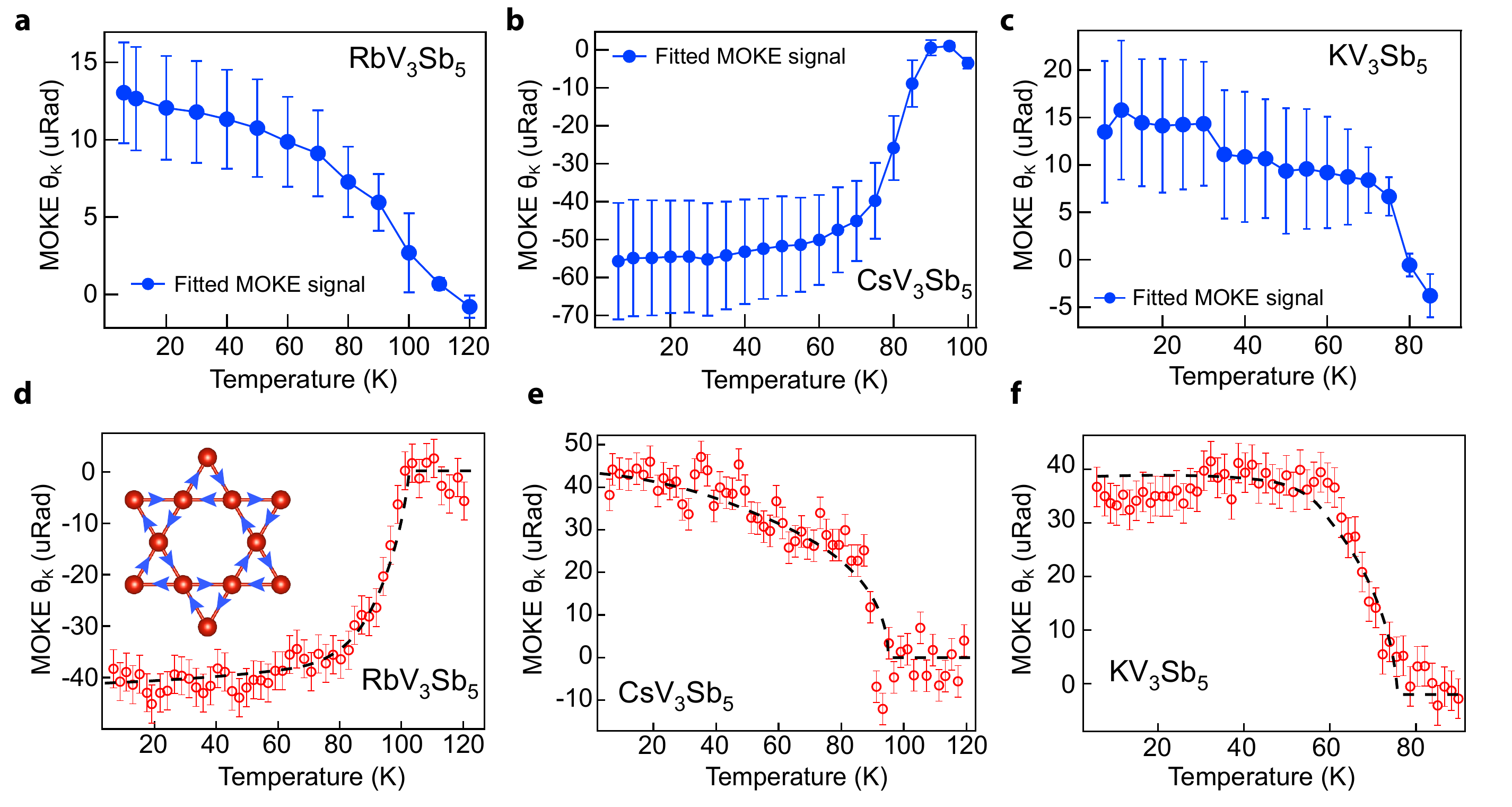}
    \caption{\textbf{MOKE in AV$_3$Sb$_5$.} (a-c) Fitted MOKE signal $\theta_K$ vs temperature with error bars for Rb, Cs and K compounds, respectively. The error bar is defined as the fitting error of $\theta_K$ by Eq. \ref{eq:eq1}. (d-f) MOKE signal vs temperature measured at the incident angles with zero birefringence for Rb, Cs and K  compounds, respectively. The inset in panel d shows the orbital currents (blue arrows) in the kagome lattice. The error bar is 3.7 uRad as defined in the main text, which is larger than the standard deviation of the $\theta_K$ from 6-40 K }
    \label{fig:fig3}
\end{figure*}

In Eq. (\ref{eq:eq1}), there is  an possible isotropic term $\theta_K$ coming from the polar MOKE due to TRS breaking. The isotropic MOKE term appears as an offset in the $\theta_T$ vs $\phi$ plot (Fig.\ref{fig:fig1}e, Fig.\ref{fig:fig2}b,d and Extended data Fig.\ref{extfig3}b, d). We extract the temperature dependent MOKE term by fitting the angle dependence by  Eq. \ref{eq:eq1}, and the results for the three compounds are plotted in Fig. \ref{fig:fig3}a-c. RbV$_3$Sb$_5$, CsV$_3$Sb$_5$, and KV$_3$Sb$_5$ show the onset of the MOKE at approximately 103 K, 92 K and 74 K, respectively. Our result agrees with another study on the Cs compound \cite{WuarXiv2021}. The fitting also yield one principal axis direction $\phi_0$. 

We use a second method to measure the MOKE with denser temperature steps by setting the incident polarization at the principal axis to eliminate the birefringence. The temperature dependent MOKE by this method for RbV$_3$Sb$_5$, CsV$_3$Sb$_5$, and KV$_3$Sb$_5$ are shown in Fig. \ref{fig:fig3}d-f. They clearly show that the onset of the MOKE signal is universally at $T_{CDW}$ for  AV$_3$Sb$_5$. There is an error bar of $\pm0.8^\circ$  in determining the principal axis, but as Extended Fig. \ref{extfig4} shows, the temperature dependent MOKE at $\phi_0\pm0.8^\circ$ still exhibit the onset of MOKE signal at $T_{CDW}$. Also, a simple estimate assuming that the exact zero birefringence angle is off by $0.8^\circ$ gives an error bar of 3.7 $\mu$rad  ($ \mathrm{sin(0.8^\circ)} \times 270$ $\mu$rad), which is much smaller than the observed MOKE signal.

Note that the two methods in Fig. \ref{fig:fig3} are performed at different locations, which lead to the different magnitude and signs. Different MOKE signs at various regions are also consistent with two TRS breaking domains. To confirm the consistency between the two methods, we measure the MOKE signal again by the second method in region 2 of RbV$_3$Sb$_5$, and obtain similar data as method 1 as shown in Extended data Fig. \ref{extfig1}d. Furthermore, as shown in Extended Fig. \ref{extfig5}, thermal cycles at the same location shown that the contour of the birefringence domain does not change much. We also find that thermal cycle is not easy to change the sign of the MOKE in all three compounds, which suggests some unknown pinning mechanism.    We would like to point out that the other possible origin of the MOKE is the spin density wave, but there has been no positive evidence of spin density wave from any measurement.  To summarize, our MOKE measurement at zero field directly probe the TRS-breaking order parameter and demonstrate that TRS is broken at the $T_{CDW}$ for all three AV$_3$Sb$_5$ compounds, indicating the existence of the long-sought loop currents in the CDW phases in AV$_3$Sb$_5$ as shown in the inset of Fig. \ref{fig:fig3}d.  \\

\textbf{Circular dichroism}

\begin{figure*}
    \centering
    \includegraphics[width=\textwidth]{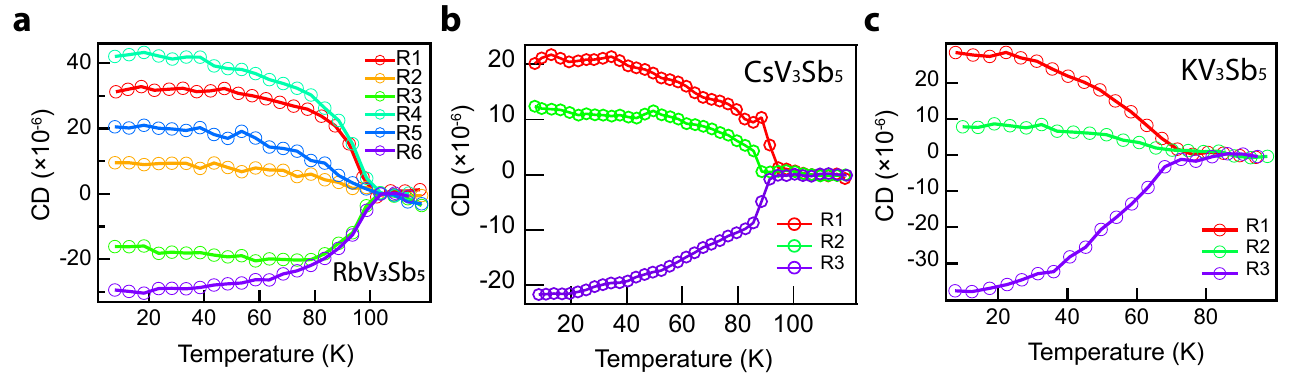}
    \caption{\textbf{Circular dichroism in AV$_3$Sb$_5$.} (a) Circular dichroism vs temperature measured at the corresponding spots in region 1-6 in Extended data Fig. \ref{extfig6}(a) in RbV$_3$Sb$_5$. (b-c) Circular dichroism vs temperature at the selected spots in region 1-3 in CsV$_3$Sb$_5$ and KV$_3$Sb$_5$ in Extended data Fig. \ref{extfig6}(b) and (c), respectively. The error bar is smaller than the symbol size of the data point in a-c. }
    \label{fig:fig4}
\end{figure*}

To further confirm TRS breaking in the CDW phase, we measure the circular dichroism (CD) on these three compounds.  Left circularly polarized  and right circularly polarized light are normally incident on the sample, and the difference of the reflectivity between left circularly polarized  and right circularly polarized light is defined as the CD. The CD measurement is free from the birefringence effect and fitting errors. The measured CD signal can be shown to be proportional to the ellipticity, $\eta$, of the MOKE contribution (see Methods).  Fig. \ref{fig:fig4} shows the CD vs temperature for all three compounds at different spatial locations, and we clearly see an onset of CD at $T_{CDW}$ as the temperature-dependent CD at different spots of the samples splits at $T_{CDW}$. The locations of these points are shown in the spatial CD mapping in Extended Fig. \ref{extfig6}. The sharp transition in $\theta_T$  is more consistent with a first order transition in the Cs sample, which is consistent with NMR/NQR measurements  \cite{Luo_npjQM_2022, Mu_CPL_2021, Song_SCP_2022}. The sharp transition is also observed  in the birefringence and MOKE measurement as shown in Extended data Fig. \ref{extfig2}a. 

Note that CD can also originate from chiral (handed) structures, and there has been theoretical proposals that the CDW on the surface of AV$_3$Sb$_5$ could be chiral if the period along the c axis is of four unit cells \cite{ParkPRB2021}. As shown in Methods, if both TRS and inversion exist, CD vanishes.  We performed second harmonic generation experiments, and observe that the SHG signal is only around 0.2 counts of photons per second under 12 mW incident power. See Extended data Fig. \ref{extfig7}. We also do not see any change across the $T_{CDW}$ in SHG. Our observation  of AV$_3$Sb$_5$ being centrosymmetric agrees with previous studies \cite{OrtizPRX2021}, and our detection sensitivity is much higher \cite{NiPRL2021, NiNatNano2021}. Therefore, the presence of CD indicates breaking TRS. What is more, in the CsV$_3$Sb$_5$, within the same birefringence domain as shown in Extended data Fig. \ref{extfig8}, we observe large-area two CD domains with opposite signs, which is another strong evidence of TRS breaking. Therefore, we conclude that the onset of CD at the $T_{CDW}$ comes from TRS breaking.     
\\

\textbf{Discussion}

 Our observation of two-fold symmetry and three-state nematicity just below $T_{CDW}$ is very different from the nematic order observed by STM in CsV$_3$Sb$_5$\cite{ZhaoNature2021,NieNature2022} and KV$_3$Sb$_5$ \cite{LiNatPhys2022}, which shows the nematic order at temperature much below $T_{CDW}$. The difference is probably not just because that STM is a surface sensitive measurement, and our probe is a bulk measurement as the penetration depth of the light is around 50 nm \cite{ZhouPRB2021,UykurNPJQuant2022}. It could be that the nematicity observed by STM is some kind of additional electronically-driven phenomena at low temperature as proposed recently\cite{ZhaoNature2021,NieNature2022,LiNatPhys2022}. Also, the three domains were not all resolved by STM\cite{ZhaoNature2021,NieNature2022,LiNatPhys2022} probably because of the large domain size. The origin of the two-fold symmetry and three domains observed in our measurement is most likely due to the $\pi$ phase shift of the stacking between CDW layers, as shown in Fig. \ref{extfig1}b and Extended data Fig.\ref{extfig1}a, because the onset temperature of birefrigence coincides with $T_{CDW}$ in all three compounds. There have been lots of works of proposing different kinds of stacking between CDW layers \cite{NieNature2022, RatcliffPRMat2021,OrtizPRX2021,MiaoPRB2021, LiPRX2021,LiangPRX2021}.  Our results shows that those without  two-fold symmetry are not compatible. The domains we observe are on the order of 100 $\mu$m scale, which explains the nematicity observed in transport experiments on macroscopic samples possibly due to unequal population of three domains \cite{XiangNatComm2021, NiCPL2021}. Our results add strong constraints on interpreting other macroscopic measurements including photo-emission, x-ray and optical spectroscopy \cite{KangNatPhys2022, LuoNatComm2022,NakayamaPRX2022, MiaoPRB2021, LiPRX2021,LiangPRX2021, OrtizPRX2021, ZhouPRB2021, UykurNPJQuant2022, NakayamaPRB2021}.

As predicted in  theoretical works \cite{ParkPRB2021, KieselPRL2013, DennerPRL2021, LinPRB2021, Christensen_PRB_2021, TanPRL2021}, the interactions between saddle points in the kagome lattice lead to various competing orders such as real and chiral flux CDW orders in AV$_3$Sb$_5$. In some parameter regimes, the favored chiral flux CDW order can also induce  real CDW order, leading to a mixture of order parameters. The results from our optical measurements shows that both the three-state nematicity and TRS breaking exhibit at T$_{CDW}$ in AV$_3$Sb$_5$, which indicate that the two kinds of symmetry breaking might be intertwined.    The TRS breaking is favorable to the chiral flux CDW order with loop currents, which might indicate unconventional superconductivity in these compounds as the superconducting phase develops from the CDW phase.   As MOKE is directly related with the transverse and longitudinal optical conductivity instead of the magnetization, it can not be used to directly estimate the effective  moment. However, it might be worthwhile to point out that the MOKE signal in AV$_3$Sb$_5$ is as small as Mn$_3$Sn\cite{Higo_NatPhotonics_2018}. The moment in Mn$_3$Sn is 0.001 $\mu_B$, so perhaps the effective moment in AV$_3$Sb$_5$ is even smaller as it was not detected by spin susceptibility and  $\mu$SR measurement. Looking forward, we hope our work will stimulate future works  to study the nematicity, the effective moment, and TRS breaking in both the CDW and superconducting phases in AV$_3$Sb$_5$. The imaging methodologies developed here can also be widely applied to other strongly correlated and topological systems. 

\section{Acknowledgement}
 We would like to thank C. Varma, Z. Wang, and I. Zeljkovic for helpful discussions. This project is mainly supported by L.W.'s startup package at the University of Pennsylvania. The development of the imaging systems was sponsored by the Army Research Office and was accomplished under Grant Number W911NF-21-1-0131, W911NF-20-2-0166, and W911NF-19-1-0342, and the Vice Provost for Research University Research Foundation. Y.X. is  also partially supported by the NSF EAGER grant  via the CMMT program (DMR-2132591), a seed grant from NSF funded Penn MRSEC  (DMR-1720530), and the Gordon and Betty Moore Foundation’s EPiQS Initiative, Grant GBMF9212 to L.W.. Z.N. acknowledges support from Vagelos Institute of Energy  Science  and  Technology  graduate  fellowship  and the Dissertation Completion Fellowship at  the  University  of  Pennsylvania.  B.R.O. and S.D.W.  acknowledge support via the UC Santa Barbara NSF Quantum Foundry funded via the Q-AMASE-i program under award DMR-1906325. Q.D. is partially supported by the NFS EPM program under Grant No. DMR-2213891.  B.Y. acknowledges funding from the European Research Council (ERC) under the European Union’s Horizon 2020 research and innovation programme (ERC Consolidator Grant ``NonlinearTopo'', No. 815869). L.B. is supported by the NSF CMMT program under Grant No. DMR-2116515. L.W. acknowledges the  support  by the Air Force Office of Scientific Research under award number FA9550-22-1-0410.  
 
\section{Author Contribution}
L.W. conceived and supervised the project.   Y.X. performed the experiments and analyzed the data with Z.N., Q.D, and L.W.. Y.L. and B.Y. performed the CD symmetry analysis. B.O. and S.W. grew the crystals. L.W.,Y.X., S.W., B.Y., and L.B. discussed and interpreted the data.      L.W. and Y.X. wrote the manuscript from input of all authors. All authors edited the manuscript.

\textit{Correspondence: }Correspondence and requests for materials
should be addressed to L.W. (liangwu@sas.upenn.edu)

\textit{Competing Interests: }The authors declare that they have no
competing financial interests.

\newpage
\setcounter{figure}{0}
\renewcommand{\figurename}{{\bf{Extended Data Figure}}}

\begin{figure*}
    \centering
    \includegraphics[width=0.9\textwidth]{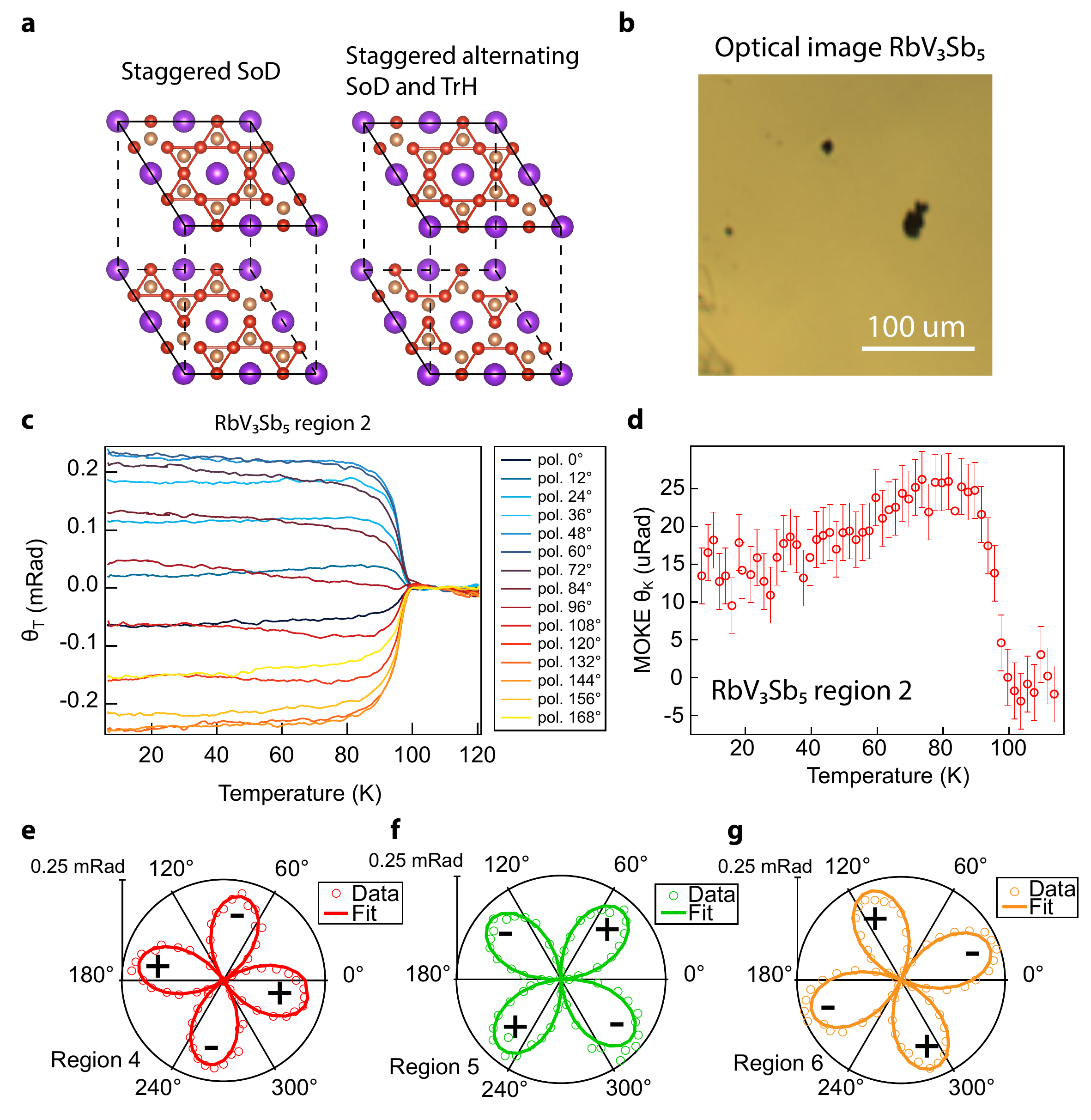}
    \caption{\textbf{Additional measurements on three-state nematic order and MOKE in RbV$_3$Sb$_5$. } (a) 3D lattice structures showing two other possible staggered CDW orders with a $\pi$ phase shift, the staggered star-of-David  (SoD) and the staggered alternating star-of-David (SoD) and tri-hexgonal (TrH) CDW orders. (b) Optical image of the mapping region in Fig. \ref{fig:fig1} (f) in RbV$_3$Sb$_5$. The black dots indicate impurities on the surface. (c) $\theta_T$ vs temperature for various incident polarization measured in region 2 in RbV$_3$Sb$_5$.  (d) MOKE signal vs temperature measured at the zero birefringence incident angle in region 2 in RbV$_3$Sb$_5$. The error bar is 3.7 uRad (see main text for definition of error). (e-g) Polar plots of the birefringence patterns at T=70 K measured at the corresponding spots in region 4, 5 and 6 (see Fig. \ref{fig:fig1}(f)), respectively. }
    \label{extfig1}
\end{figure*}

\begin{figure*}
    \centering
    \includegraphics[width=0.9\textwidth]{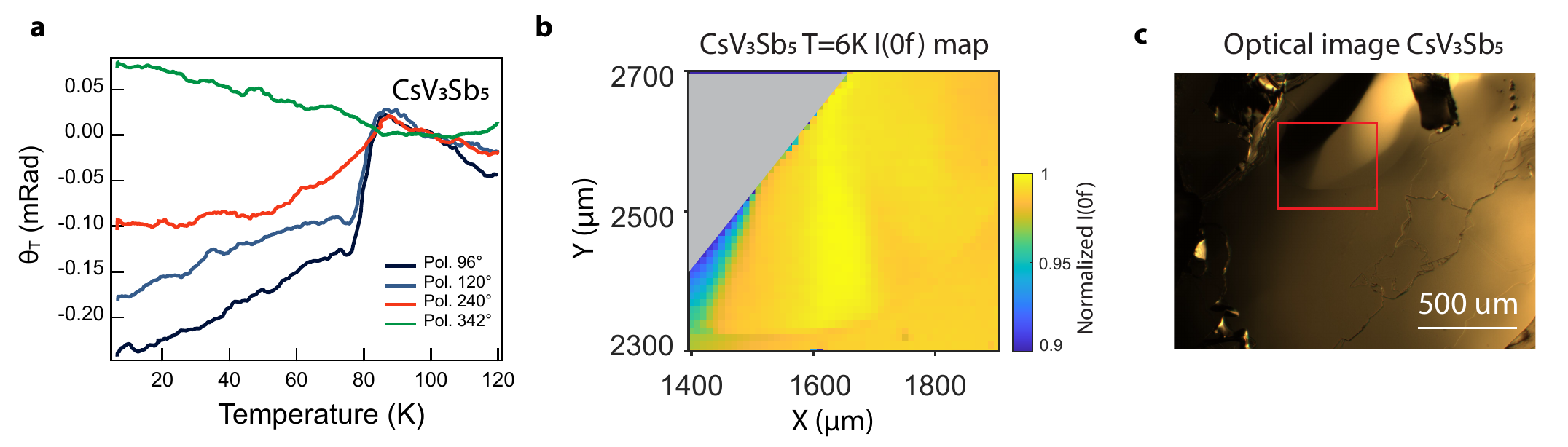}
    \caption{\textbf{$\theta_T$ vs temperature and characterization of the Cs sample in the main text.} (a)  $\theta_T$ vs temperature for various incident polarization for  the Cs sample shown in (b-c) in this figure, which is also the same sample for Fig. 2-4 in the main text and Extended data Fig.6-8.  The sharp transition in $\theta_T$ at certain polarization is more consistent with a first order transition in the Cs sample, which is consistent with NMR/NQR measurements.  Note that the Cs sample in extended data Fig.3 is a different sample, which shows a smoother transition.     (b) Mapping of the normalized I(0f) signal, the reflectivity,  in the CsV$_3$Sb$_5$ sample. A variation of the I(0f) signal is observed in the mapping data, which indicates an uneven surface. (c) Optical image of the cleaved Cs sample, the red box indicates the mapping region in (b). }
    \label{extfig2}
\end{figure*}

\begin{figure*}
    \centering
    \includegraphics[width=\textwidth]{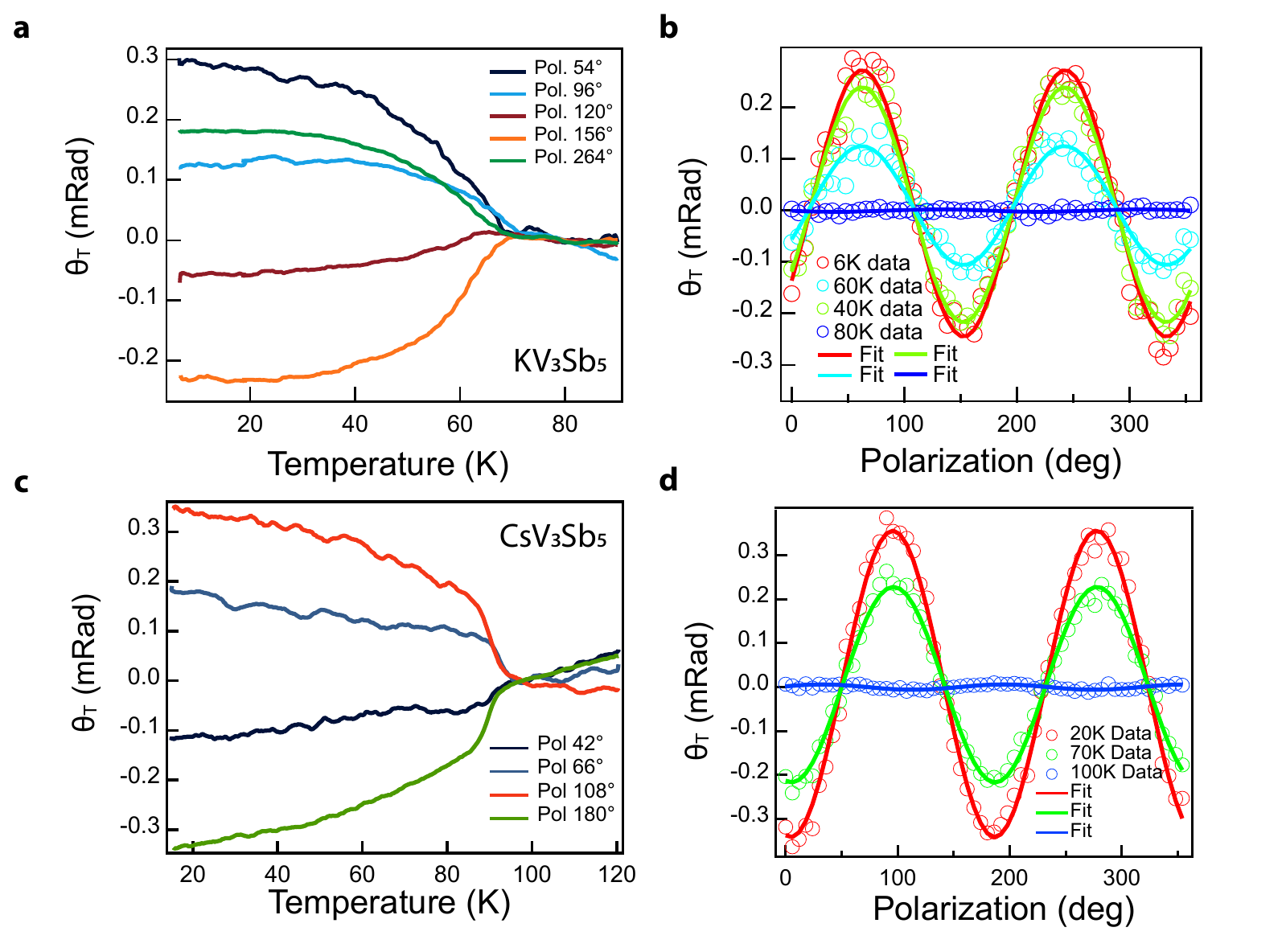}
    \caption{\textbf{Dependence of $\theta_T$ on temperature and incident polarization of K and Cs compound.} (a, c) $\theta_T$ vs temperature for various incident polarization for KV$_3$Sb$_5$ and CsV$_3$Sb$_5$, respectively. (b, d) $\theta_T$ vs incident polarization at different temperature cuts for KV$_3$Sb$_5$ and CsV$_3$Sb$_5$, respectively. }
    \label{extfig3}
\end{figure*}

\begin{figure*}
    \centering
    \includegraphics[width=\textwidth]{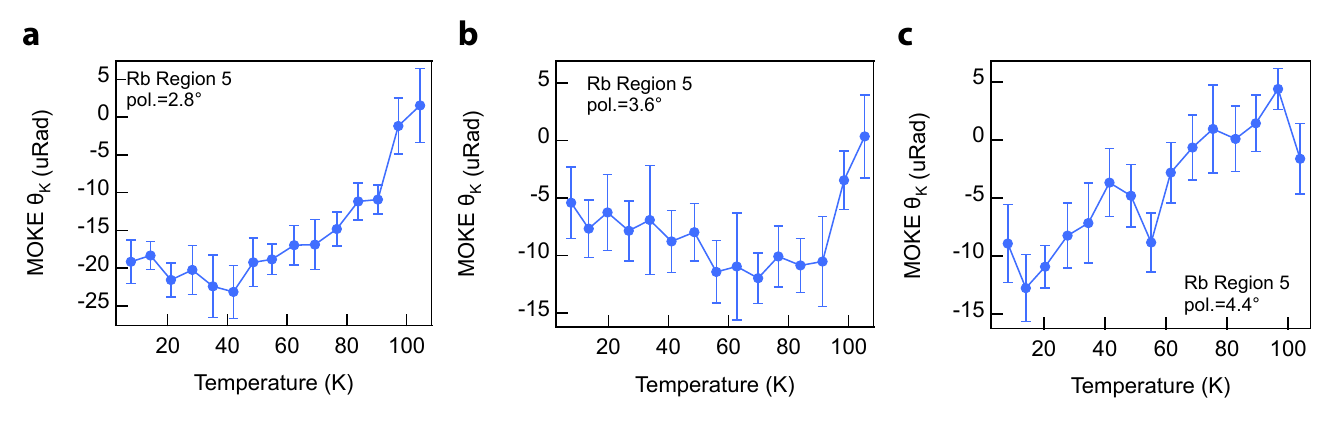}
    \caption{\textbf{Temperature dependent MOKE at  $\phi_{0}$ and $\phi_{0}\pm 0.8$.} (a-c) MOKE signal measured at the incident angles $\phi_{0}=3.6^\circ$ and $\phi_{0}\pm 0.8$ in region 5 for RbV$_3$Sb$_5$, where $\phi_{0}$ is the incident angle that birefringence contribution is zero. The error bar is defined as the statistical error for data points averaged together over 2 K range bins. }
    \label{extfig4}
\end{figure*}

\begin{figure*}
    \centering
    \includegraphics[width=0.8\textwidth]{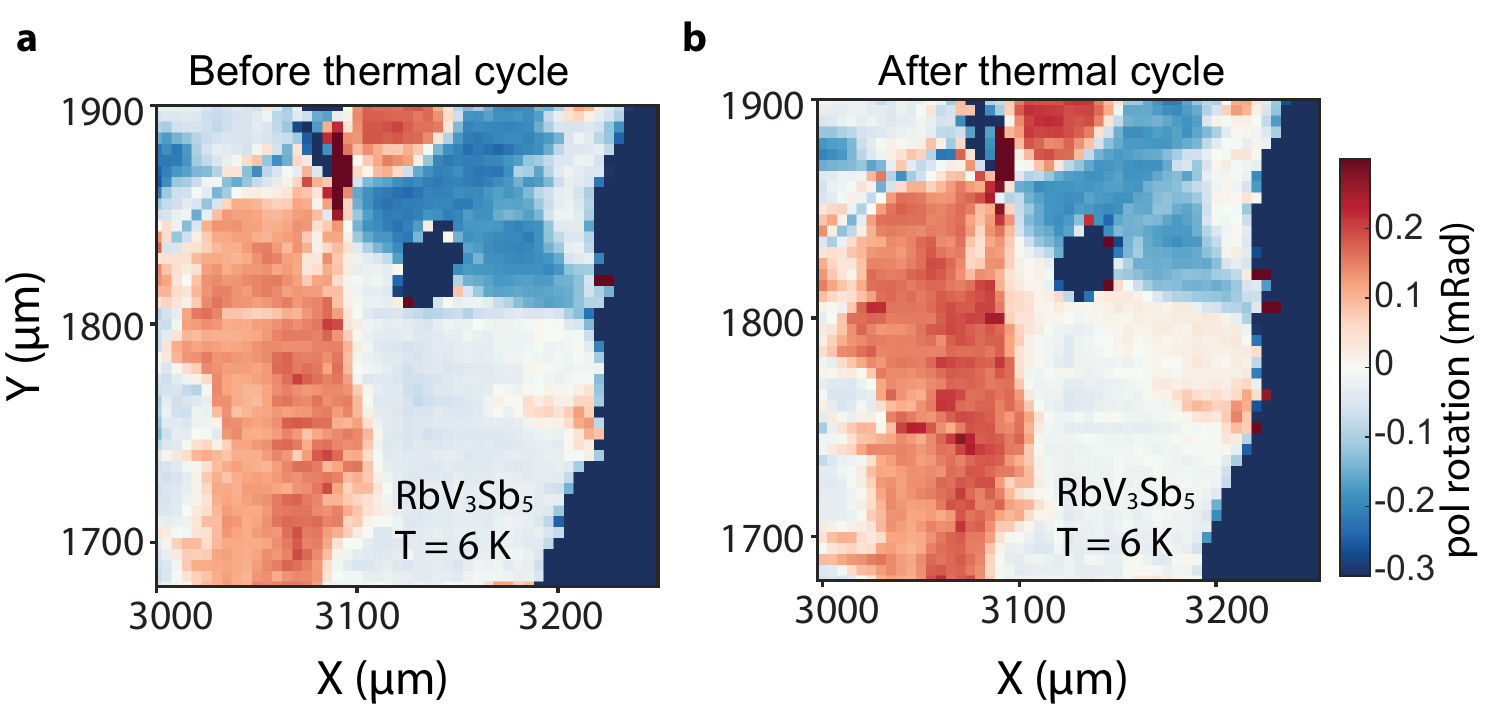}
    \caption{\textbf{Birefringence domains under thermal cycles.} (a-b) Spatial mapping of $\theta_{T}$ at T = 6 K, before and after thermal cycles for RbV$_3$Sb$_5$.}
    \label{extfig5}
\end{figure*}

\begin{figure*}
    \centering
    \includegraphics[width=\textwidth]{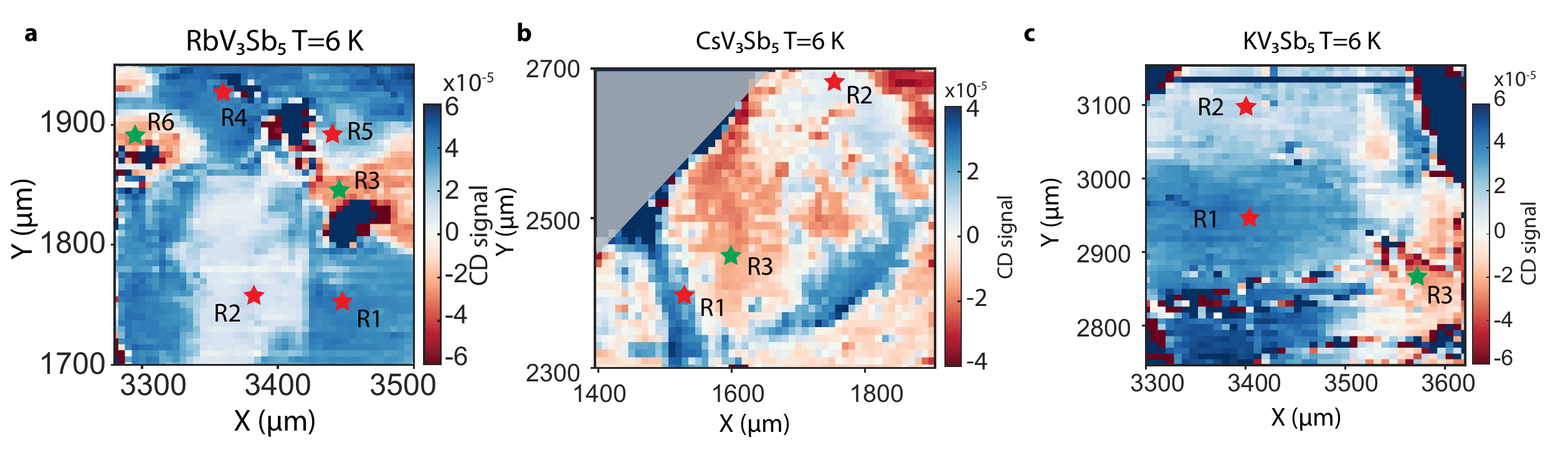}
    \caption{ \textbf{Circular dichroism maps of AV$_3$Sb$_5$.} (a-c) Circular dichroism maps at T= 6 K for Rb, Cs and K compounds, respectively. The red and green star symbols indicate the positions where circular dichroism vs temperature measurements are performed in Fig. \ref{fig:fig4}.    }
    \label{extfig6}
\end{figure*}

\begin{figure*}
    \centering
    \includegraphics[width=\textwidth]{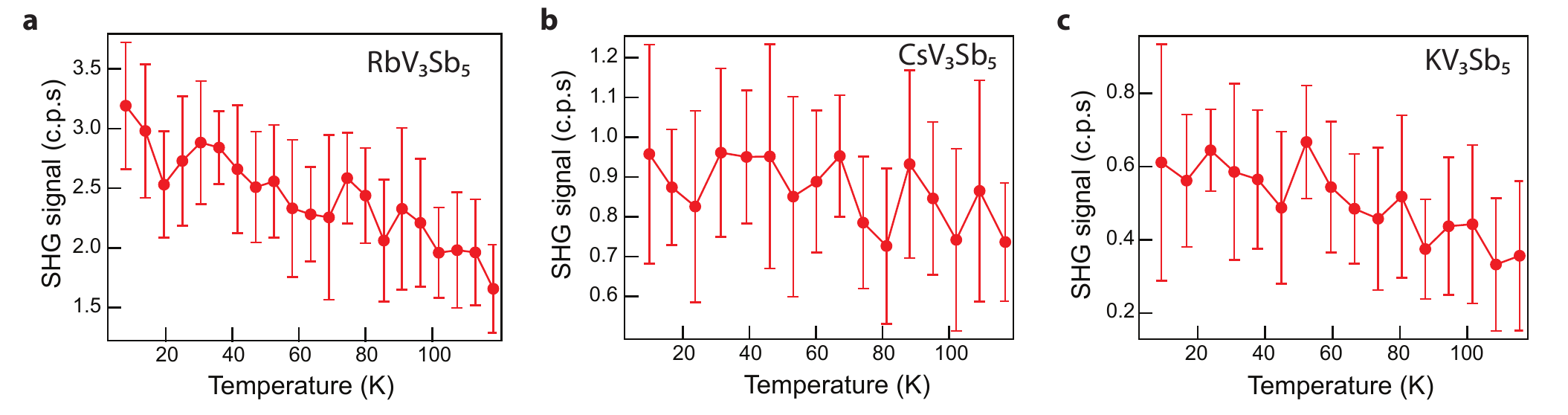}
    \caption{\textbf{Second harmonic generation  of AV$_3$Sb$_5$.} (a-c) Second harmonic generation vs temperature for Rb, Cs and K compounds, respectively. The error bar is defined as the statistical error for data points averaged together over 5 K range bins. }
    \label{extfig7}
\end{figure*}

\begin{figure*}
    \centering
    \includegraphics[width=0.85\textwidth]{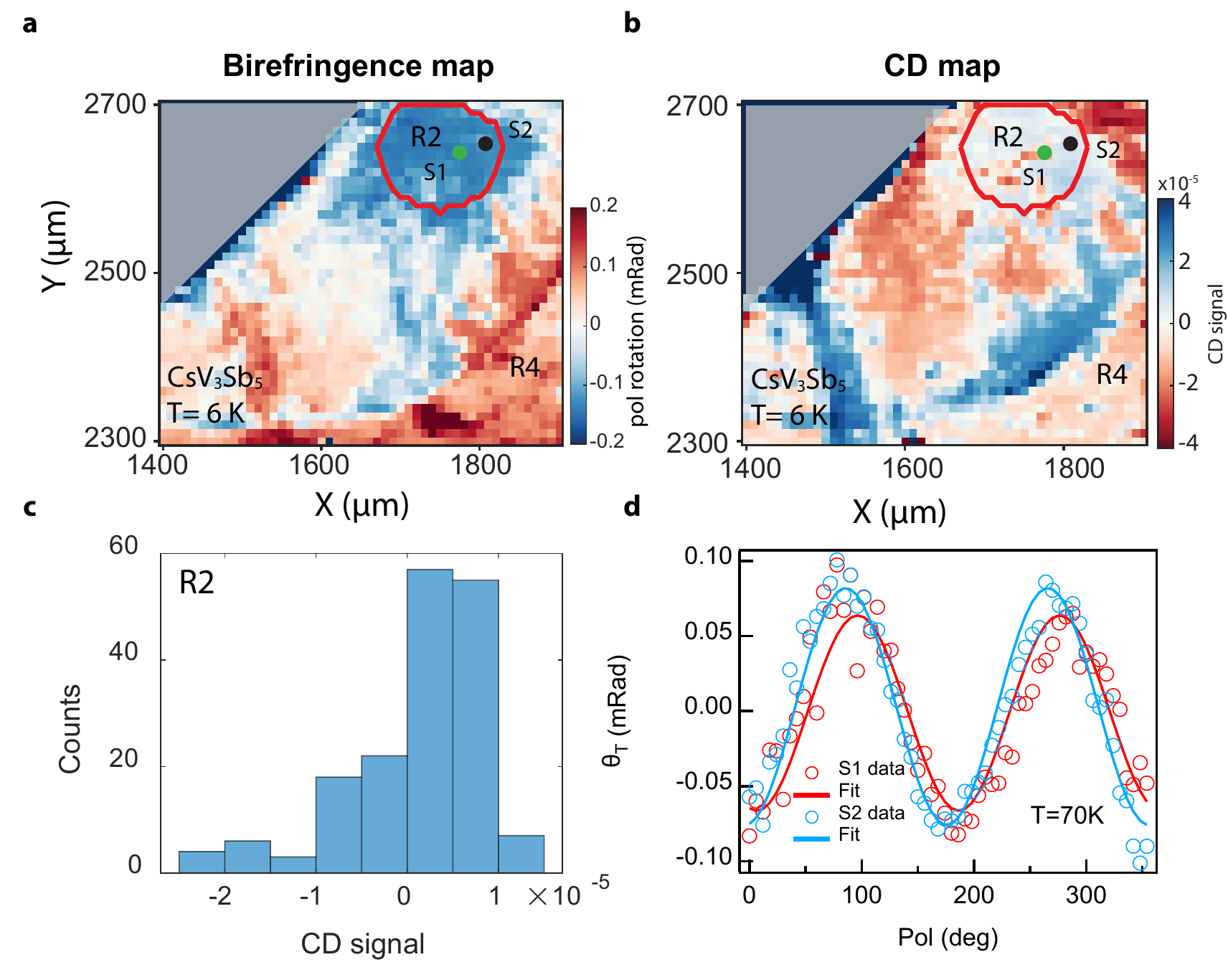}
    \caption{\textbf{Comparison of birefringence and circular dichroism maps of CsV3Sb5.} (a-b) The spatial birefringence and circular dichroism maps of CsV$_3$Sb$_5$ at T = 6 K, respectively. The region circled by red is region 2, where points within each region have the same birefringence pattern. The green and black dots indicate two spots in R2 (S1 and S2), which have same birefringence patterns but opposite signs of CD signal. (c) Histograms of circular dichroism signals within region 2. Both positive and negative CD signals exist within R2. (d) $\theta_T$ vs incident polarization at T= 70 K measured at S1 and S2 within R2, respectively.}
    \label{extfig8}
\end{figure*}

\begin{figure*}
    \centering
    \includegraphics[width=0.9\textwidth]{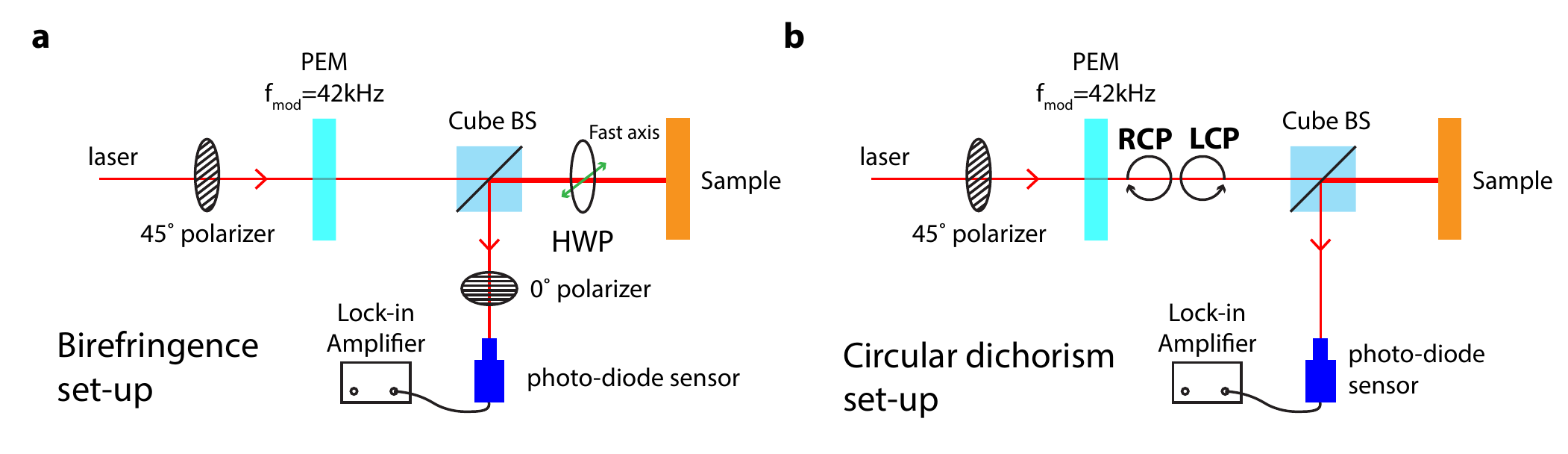}
    \caption{\textbf{Optical set-ups in this paper.} Optical set-ups for the birefringence (a) and circular dichroism (b) measurements.}
    \label{extfig9}
\end{figure*}

\clearpage
\newpage

\section{Methods}

\subsection{Sample growth method}
Single crystals of CsV$_3$Sb$_5$, RbV$_3$Sb$_5$, and KV$_3$Sb$_5$ were synthesized from Cs (liquid, Alfa 99.98\%), Rb (liquid, Alfa 99.75\%), K (metal, Alfa 99.95\%), V (powder, Sigma 99.9\%) and Sb (shot, Alfa 99.999\%). As-received vanadium powder was purified in-house to remove residual oxides. Due to extreme reactivity of elemental alkalis, all further preparation of AV$_3$Sb$_5$ crystals were performed in an argon glovebox with oxygen and moisture levels $<$0.5\,ppm. AV$_3$Sb$_5$ single crystals were synthesized using the self-flux method. Elemental reagents were milled in a pre-seasoned tungsten carbide vial to form a composition which is 50\,at.\% A$_{0.4}$Sb$_{0.6}$ and approximately 50\,at.\% VSb$_2$. Excess antimony can be added to the flux to suppress volatility if needed. The fluxes were loaded into alumina crucibles and sealed within stainless steel jackets. The samples were heated to 1000 $^{\circ}$ C at 250 $^{\circ}$ C/hr and soaked there for 24\,h. The samples were subsequently cooled to 900 $^{\circ}$ C at 100 $^{\circ}$ C/hr and then further to 500 $^{\circ}$ C at 1 $^{\circ}$ C/hr. Once cooled, the crystals are recovered mechanically. Crystals are hexagonal flakes with brilliant metallic luster. Elemental composition of the crystals was assessed using energy dispersive x-ray spectroscopy (EDX) using a APREOC scanning electron microscope.

\subsection{Birefringence and MOKE measurements}
Laser pulses from a Ti:sapphire oscillator with 800 nm center wavelength, 80 MHz repetition rate and 50 fs pulse duration are used to measure the change of polarization angle.   To measure the change of the polarization angle without rotating the sample, a half wave plate (HWP) is put right in front of the sample and in between the cube beam splitter. Both the incident and reflected light goes through the half wave plate, thus polarization change due to the half wave plate is cancelled. The net change of the polarization purely comes from the sample itself. By rotating the HWP, we can measure the change of polarization $\theta_T$ for different incident polarizations. The background from the HWP is calibrated at temperature around 20 K above the $T_{CDW}$. The optical set-up for the birefringence measurement is shown in Extended data Fig.\ref{extfig9}(a). All of the scanning imaging experiments in this work are performed by fixing the laser beam spot but moving the samples by Attocube xyz positioners.

In the following derivation, we will show how the birefringence and the MOKE signal can be distinguished by rotating the half wave plate and keeping the beam at the same spot. 
The Jones matrix for polarizer at 45$^\circ$ and 0$^\circ$, photo-elastic modulator (PEM), half wave plate ($\alpha$) and mirror are,
\begin{align*}
    P(45)&= \frac{1}{2}\begin{bmatrix}
    1 & 1\\
    1 & 1
    \end{bmatrix}\\
    P(0) & =\begin{bmatrix}
    1 & 0\\
    0 & 0
    \end{bmatrix}\\
    HWP(\alpha) & = \begin{bmatrix}
    \cos(2\alpha) & \sin(2\alpha)\\
    \sin(2\alpha) & -\cos(2\alpha)
    \end{bmatrix}\\
    PEM &= \begin{bmatrix}
    1 & 0\\
    0 & e^{i\tau}
    \end{bmatrix}\\
    M&= \begin{bmatrix}
    1 & 0\\
    0 & -1
    \end{bmatrix}
\end{align*}
where $\alpha$ is the angle between the fast axis of the half wave plate and the horizontal axis of the lab, $\tau$ is the phase retardation applied by the PEM. The rotation matrix is,
\begin{equation*}
    R(\beta)=\begin{bmatrix}
    \cos(\beta) & -\sin(\beta)\\
    \sin(\beta) & \cos(\beta)
    \end{bmatrix}
\end{equation*}
where $\beta$ is the angle of rotation. For a sample that has both the birefringence and MOKE effect, we can use the following Jones matrix to represent the sample (ignoring the higher order term),
\begin{equation}
    S(\theta)=\begin{bmatrix}
    1-\sin(\theta)^2\Delta  &  \frac{\Delta\sin(2\theta)}{2} -c\\
     \frac{\Delta\sin(2\theta)}{2}+c & 1-\cos(\theta)^2 \Delta
    \end{bmatrix}
\end{equation}
where $\theta$ is the angle between the fast axis and the polarization of the incident light, $\Delta=\delta + i\kappa$ is the complex birefringence term, and $c=\theta_K+i\eta$ comes from the MOKE effect which is also a complex number. The output light $O$ measured at the photo detector can be calculated by,
\begin{align*}
    O&=P(0) \cdot HWP(\pi-\alpha) \cdot M \cdot  S(\theta) \cdot HWP(\alpha) \\
    &\cdot PEM \cdot P(45) \cdot \begin{bmatrix} E \\0
    \end{bmatrix}\\
    &=E\begin{bmatrix}
    \frac{\Delta \,\mathrm{cos}\left(4\,\alpha -2\,\theta \right)}{4}+\frac{c\,{\mathrm{e}}^{\tau \,\mathrm{i}} }{2}-\frac{\Delta }{4}+\frac{\Delta \,{\mathrm{e}}^{\tau \,\mathrm{i}} \,\mathrm{sin}\left(4\,\alpha -2\,\theta \right)}{4}+\frac{1}{2}\\0
    \end{bmatrix}
\end{align*}
We see that the above expression only depends on the angle difference term $4\alpha-2\theta$, thus we can set $\theta=0$ and it will not affect the final results. The intensity measured at the photo detector is (ignoring the higher order terms),
\begin{align*}
    I(t)&=|O|^2=E^2\bigg[  \mathrm{sin}\left(\tau \right)\,{\left(-\frac{\eta }{2}-\frac{\kappa \,\mathrm{sin}\left(4\,\alpha \right)}{4}\right)} \\&+\mathrm{cos}\left(\tau \right)\,{\left(\frac{\theta_{K} }{2}+\frac{\delta \,\mathrm{sin}\left(4\,\alpha \right)}{4}\right)}+ \frac{\delta \,\mathrm{cos}\left(4\,\alpha \right)}{4}-\frac{\delta }{4}
    \\&+\frac{1}{4} \bigg] +  \mathcal{O}(h)
\end{align*}
Setting $\tau=\tau_{0}\sin(\omega t)$, and using the Fourier decomposition of $\cos(\tau(t))$ and $\sin(\tau(t))$, we have the following relation,
\begin{align*}
    I(t)&=E^2\bigg[  \frac{\delta \,\mathrm{cos}\left(4\,\alpha \right)}{4}-\frac{\delta }{4}+\frac{1}{4}\\
    &+2J_1(\tau_0)\sin(\omega t){\left(-\frac{\eta }{2}-\frac{\kappa \,\mathrm{sin}\left(4\,\alpha \right)}{4}\right)}\\ &+(J_0(\tau_0)+2J_2(\tau_0)\cos(2\omega t)){\left(\frac{\theta_{K}}{2}+\frac{\delta \,\mathrm{sin}\left(4\,\alpha \right)}{4}\right)}\bigg] 
    \\&+  \mathcal{O}(h)
\end{align*}
The $1f, 2f$ and DC component of the signal is,
\begin{align*}
    I(1f)&= E^2 2 J_1(\tau_0) {\left(-\frac{\eta }{2}-\frac{\kappa \,\mathrm{sin}\left(4\,\alpha \right)}{4}\right)} \sin(\omega t)\\
    I(2f)&= E^2 2 J_2(\tau_0) {\left(\frac{\theta_{K} }{2}+\frac{\delta \,\mathrm{sin}\left(4\,\alpha \right)}{4}\right)} \cos(2\omega t)\\
    I(DC)&= E^2 \Big[ \frac{\delta \,\mathrm{cos}\left(4\,\alpha \right)}{4}-\frac{\delta }{4}+\frac{1}{4}
    \\&+ J_0(\tau_0) (\frac{\theta_{K} }{2}+\frac{\delta \,\mathrm{sin}\left(4\,\alpha \right)}{4}) \Big]
\end{align*}
We can set $\tau_0=2.405$, which is the zero point for the $J_0$ Bessel function. Since both $\delta$ and $\theta_{K}$ are very small, we can approximate the DC term by $I(DC)=E^2/4$. Furthermore, lock-in measures the RMS (root-mean-square) of the signal, so we have the following relation,
\begin{align}
    \frac{I_{lock}(1f)}{I_{lock}(DC)}&=\frac{4J_1(\tau_0)}{\sqrt{2}}{\left(-\eta -\frac{\kappa \,\mathrm{sin}\left(4\,\alpha \right)}{2}\right)}\\
    \frac{I_{lock}(2f)}{I_{lock}(DC)}&=\frac{4J_2(\tau_0)}{\sqrt{2}}{\left(\theta_{K} +\frac{\delta \,\mathrm{sin}\left(4\,\alpha \right)}{2}\right)}
\end{align}
Finally, we note that the polarization $\phi$ of the incident light changes twice as much as the change of the half wave plate angle $\phi=2\alpha$, we have the following relation,
\begin{align}
    -\eta -\frac{\kappa \,\mathrm{sin}\left(2\,\phi \right)}{2}&= \frac{\sqrt{2}}{4J_1(\tau_0)}\frac{I_{lock}(1f)}{I_{lock}(DC)}\\
    \theta_{K} +\frac{\delta \,\mathrm{sin}\left(2\,\phi \right)}{2}&=\frac{\sqrt{2}}{4J_2(\tau_0)}\frac{I_{lock}(2f)}{I_{lock}(DC)} \label{eq:bire_angle}
\end{align}
We can see that the change of the polarization angle in Eq. (\ref{eq:bire_angle}) has a constant MOKE term $\theta_{K}$, and a $\phi$ dependent birefringence term $\delta \,\mathrm{sin}(2\phi)/2$.

\subsection{CD measurement}
The optical set-up for the circular dichroism measurement is shown in Extended Fig.\ref{extfig9}(b). An alternating left circularly polarized  and right circularly polarized  light generated by the photo elastic modular (PEM) is reflected off the sample, and the difference between the left circularly polarized and right circularly polarized intensity (CD) is measured by a photo detector connected to a lock-in analyzer. PEM modulates the incident light between the left circularly- and right circularly- polarized at a frequency of $f=$42 kHz. Both the $1f$ and DC component is extracted from the measured signal, and the circular dichroism of the sample is proportional to $I(1f)/I(DC)$.

Similar to the previous derivation, we can write out the Jones matrix for circular dichroism. The output light after going through a 45 degree polarizer, PEM and sample is,
\begin{align*}
    O&=M\cdot S(\theta) \cdot PEM \cdot P(45) \cdot \begin{bmatrix}
    E\\
    0
    \end{bmatrix}\\
    &=E\begin{bmatrix}
    -\frac{\Delta \,{\mathrm{sin}\left(\theta \right)}^2 }{2}-\frac{{\mathrm{e}}^{\tau \,\mathrm{i}} \,{\left(c-\frac{\Delta \,\mathrm{sin}\left(2\,\theta \right)}{2}\right)}}{2}+\frac{1}{2}\\
    -\frac{c}{2}+\frac{{\mathrm{e}}^{\tau \,\mathrm{i}} \,{\left(\Delta \,{\mathrm{cos}\left(\theta \right)}^2 -1\right)}}{2}-\frac{\Delta \,\mathrm{sin}\left(2\,\theta \right)}{4}
    \end{bmatrix}
\end{align*}
The intensity measured by the photo detector is,
\begin{align*}
    I(t)&=E^2\left( \mathrm{sin}\left(\tau \right)\eta  +\mathrm{cos}\left(\tau \right)\,\frac{\delta \,\mathrm{sin}\left(2\,\theta \right)}{2}-\frac{\delta}{2}+\frac{1}{2}\right) 
    \\&+ \mathcal{O}(h)\\
    &=E^2\bigg[\frac{1}{2}-\frac{\delta}{2}+ 2J_1(\tau_0)\mathrm{sin}(\omega t) \eta \\&+\Big(J_0(\tau_0)+2J_2(\tau_0)\mathrm{cos}(2\omega t)\Big)\frac{\delta \,\mathrm{sin}\left(2\,\theta \right)}{2} \bigg]\\
    &+\mathcal{O}(h)
\end{align*}
where $\tau_0=\frac{\pi}{2}$. For circular dichroism $I_{CD}$, we are measuring the ratio between the $1f$ and the $DC$ component which is equal to,
\begin{equation}
    I_{CD}=\frac{I_{lock}(1f)}{I_{lock}(DC)}=2\sqrt{2}J_{1}(\frac{\pi}{2}) \eta \label{eq:cd_signal}
\end{equation}
We see from Eq. (\ref{eq:cd_signal}) that the circular dichroism signal does not depend on the sample orientation ($\theta$) as expected, and is only related to the ellipticity $\eta$ term from the MOKE effect.

In our optical setup, we used a 50:50 non-polarizing cube in between the PEM and objective to collect the signal. The non-polarizing cube has different transmission coefficients for the \emph{s} and \emph{p} polarized light across the 750 - 850 nm wavelength range of our pulsed laser. To account for the this \emph{s} and \emph{p} transmission differences and rule out its effect on the circular dichroism signal, we can add in the Jones matrix for the cube $C$ into the CD derivation,
\begin{equation*}
    C=\begin{bmatrix}
    \sqrt{T_p} & 0 \\
    0 & \sqrt{T_s}
    \end{bmatrix}
\end{equation*}
where $T_s$ and $T_p$ are the transmission coefficient for the $\emph{s}$ and $\emph{p}$ polarized light. The output light in this case is,
\begin{align*}
    O&=\begin{bmatrix}
    \sqrt{1-T_p} & 0 \\
    0 & \sqrt{1-T_s}
    \end{bmatrix} \cdot M \cdot S(\theta) \cdot \begin{bmatrix}
    \sqrt{T_p} & 0 \\
    0 & \sqrt{T_s}
    \end{bmatrix}\\
    &\cdot PEM \cdot P(45) \cdot \begin{bmatrix}
    E\\
    0
    \end{bmatrix}
\end{align*}

The intensity measured by the photo detector is,
\begin{align*}
    I(t) &= -2J_1(\tau_0)\sin{(\omega t)}\sqrt{T_p T_s} \Big[ \frac{(T_s+T_p-2)\eta}{2}\\
    &+ \frac{(T_s-T_p)\kappa \sin(2\theta) }{4} \Big] - \Big(J_0(\tau_0)+2J_2(\tau_0)\cos(\omega t)\Big) \\
    &\sqrt{T_p T_s}\Big[\frac{(T_s-T_p)\theta_{K}}{2}+\frac{(T_p+T_s-2)\delta \sin(2\theta)}{4} \Big]\\
    &+\frac{T_s+T_p-T_s^2-T_p^2}{4}+\frac{\delta}{2}\Big[T_s^2-T_s+T_s\sin^2{(\theta)}\\
    &-T_p\sin^2{(\theta)}-T_s^2\sin^2{(\theta)}+T_p^2\sin{(\theta)}\Big]+\mathcal{O}(h)
\end{align*}
where $\tau_0=\frac{\pi}{2}$. Now, the circular dichroism signal becomes,
\begin{align*}
    I_{CD}&=\frac{I_{lock}(1f)}{I_{lock}(DC)}=\frac{J_1(\tau_0)\sqrt{2T_p T_s}}{T_s+T_p-T_s^2-T_p^2}\\
    &\Big[ 2(T_s+T_p-2)\eta +(T_s-T_p)\kappa \sin{(2\theta)}\Big]
\end{align*}
We see now there is a birefringence correction term in the circular dichroism signal as follows,
\begin{equation}
    I_{bire}=\frac{J_1(\eta_0)\sqrt{2T_p T_s}(T_s-T_p)\kappa \sin{(2\theta)}}{T_s+T_p-T_s^2-T_p^2}
\end{equation}
This correction term is proportional to the product of $\kappa$ and the difference in the transmission coefficients $T_s-T_p$. Given the specification of the non-polarized cube $|T_s-T_p| \sim 10^{-2}$, we can estimate the magnitude of such correction to be,
\begin{equation}
    I_{bire} \propto \kappa \cdot (T_s - T_p)= 10^{-4} \cdot 10^{-2}=10^{-6}
\end{equation}
The birefringence correction is one order of magnitude less than the circular dichroism signal of the samples, thus can be ignored in our measurement.

\subsection{CD Theory}
We derive the general CD effect and discuss consequences of TRS-breaking and crystal-symmetry-breaking on CD. According to Fermi's golden rule, the absorption rate of circular light with helicity $\sigma_\pm$ is 
\begin{equation}
    I(\sigma_\pm) = \frac{2\pi}{\hbar} \sum_{c,v} \left| \langle c| H' |v\rangle \right|^2 \delta(E_c - E_v - \hbar\omega),
\end{equation}
where $v$ and $c$ refer to valance and conduction band states, respectively, and $\hbar\omega$ is the photon energy. $H'$ is the interaction Hamiltonian, whose expression is given by
\begin{equation}\label{H'}
    H' = -e \bm{E} \cdot \bm{r} - \bm{m} \cdot \bm{B}.
\end{equation}
The two terms on the right-hand side of Eq. \eqref{H'} correspond to electric and magnetic dipole interactions, respectively. The magnetic dipole term is usually much weaker than the electric dipole term. However, the magnetic dipole cannot be ignored for CD in time-reversal invariant systems where the electric dipole contribution vanishes as illustrated following. 

For helical photon $\sigma_\pm$, the electric and magnetic field is 
\begin{equation}
    \bm{E} = E_0 (1, \pm i, 0) ~~~~~,~~~~~ \bm{B} = \frac{E_0}{c} (\mp i, 1,0),
\end{equation}
so that the interaction matrix element is 
\begin{equation}
    |\langle c| H'_\pm |v\rangle|^2 = |E_0|^2 \Big|\Big\langle c\Big| er_\pm \mp i \frac{m_\pm}{c} \Big|v\Big\rangle \Big|^2
\end{equation}
where $r_\pm = x \pm iy$, and $m_\pm = m_x \pm i m_y$. Therefore the CD is
\begin{equation}
\begin{split}\label{CD}
    & I(\sigma_+) - I(\sigma_-) \propto |\langle c|H'_+|v\rangle|^2 - |\langle c|H'_-|v\rangle|^2 \\
    = & 2\text{Im} \Bigg[ \underbrace{e^2\langle v|x|c\rangle \langle c|y|v\rangle}_{\text{\ding{172}}} \\
    & + \underbrace{\frac{e}{c} \langle v|m_x|c\rangle \langle c|x|v\rangle + \frac{e}{c} \langle v|m_y|c\rangle \langle c|y|v\rangle}_{\text{\ding{173}}} \\
    & + \underbrace{\frac{1}{c^2} \langle v|m_y|c\rangle \langle c|m_x|v\rangle}_{\text{\ding{174}}} \Bigg] .\\
\end{split}
\end{equation}

In Eq. \eqref{CD}, \ding{172} comes from electric dipole, \ding{173} originates in both electric and magnetic dipoles, and \ding{174} come from merely magnetic dipole interactions. The \ding{174} term is usually ignored since it is too small compared to \ding{172} and \ding{173}.

\ding{172} is the Berry curvature between valance band $v$ and conduction band $c$. It is even under inversion symmetry but odd under TRS. It is usually called magnetic CD in literature. \ding{173} is even under TRS but odd for mirror or inversion, because, for example, $\bra{v}m_{x}\ket{c}\bra{c}x\ket{v}$ is odd under $x/y\rightarrow -x/-y$ reflection or $(x,y,z)\rightarrow (-x,-y,-z)$ inversion. 
Therefore, if TRS is broken, both \ding{172} and \ding{173} are nonzero. If TRS is conserved, only \ding{173} can appear in a chiral material. If both TRS and inversion symmetry exist, the CD effect vanishes. In the case of inversion and/or mirror symmetries in the material, the CD effect due to \ding{172} indicates the TRS-breaking.\\

\textit{Data availability:} All data needed to evaluate the conclusions in the paper are present in the paper and the Extended Data figures. Additional data related to this paper could be requested from the authors.

\end{document}